\DocumentMetadata{}
\documentclass[sigconf]{acmart}  
\copyrightyear{2024}
\acmYear{2024}
\setcopyright{acmlicensed}
\usepackage{hyperref}       
\acmConference[LLM4Eval \@ SIGIR 2024]{Proceedings of the 47th ACM SIGIR Conference on Research and Development in Information Retrieval}{July 14--18, 2024}{Washington D.C., USA}
\acmBooktitle{Proceedings of the 47th ACM SIGIR Conference on Research and Development in Information Retrieval (SIGIR '24), July 14--18, 2024, Washington D.C., USA}

\AtBeginDocument{%
  \providecommand\BibTeX{{%
    \normalfont B\kern-0.5em{\scshape i\kern-0.25em b}\kern-0.8em\TeX}}}

\usepackage{hyperref}       
\usepackage{url}            
\usepackage{nicefrac}       
\usepackage{microtype}      
\usepackage{algorithm}  
\usepackage{algpseudocode}
\usepackage{graphicx}
\usepackage{enumitem}
\usepackage{multirow}
\usepackage{geometry}
\usepackage{float}

\begin{document}

\title{Large Language Models for Relevance Judgment\\in Product Search}

\author{Navid Mehrdad}
\affiliation{%
  \institution{Walmart Global Technology}
  \city{Sunnyvale}
  \country{USA}
}
\email{navid.mehrdad@walmart.com}

\author{Hrushikesh Mohapatra}
\affiliation{%
  \institution{Walmart Global Technology}
  \city{Sunnyvale}
  \country{USA}
}
\email{hrushikesh.mohapatra@walmart.com}

\author{Mossaab Bagdouri}
\affiliation{%
  \institution{Walmart Global Technology}
  \city{Sunnyvale}
  \country{USA}
}
\email{mossaab.bagdouri@walmart.com}

\author{Prijith Chandran}
\affiliation{%
  \institution{Walmart Global Technology}
  \city{Sunnyvale}
  \country{USA}
}
\email{prijith.chandran@walmart.com}

\author{Alessandro Magnani}
\affiliation{%
  \institution{Walmart Global Technology}
  \city{Sunnyvale}
  \country{USA}
}
\email{alessandro.magnani@walmart.com}

\author{Xunfan Cai}
\affiliation{%
  \institution{Walmart Global Technology}
  \city{Sunnyvale}
  \country{USA}
}
\email{xunfan.cai@walmart.com}

\author{Ajit Puthenputhussery}
\affiliation{%
  \institution{Walmart Global Technology}
  \city{Sunnyvale}
  \country{USA}
}
\email{ajit.puthenputhussery@walmart.com}

\author{Sachin Yadav}
\affiliation{%
  \institution{Walmart Global Technology}
  \city{Bangalore}
  \country{India}
}
\email{Sachin.Yadav1@walmart.com}

\author{Tony Lee}
\affiliation{%
  \institution{Walmart Global Technology}
  \city{Sunnyvale}
  \country{USA}
}
\email{tony.lee@walmart.com}

\author{ChengXiang Zhai}
 \affiliation{%
   \institution{UIUC}
   \city{Urbana-Champaign}
   \country{USA}
 }
\email{czhai@illinois.edu}

\author{Ciya Liao}
\affiliation{%
  \institution{Walmart Global Technology}
  \city{Sunnyvale}
  \country{USA}
}
\email{ciya.liao@walmart.com}

\renewcommand{\shortauthors}{Mehrdad et al.}

\begin{abstract}
High relevance of retrieved and re-ranked items to the search query is the cornerstone of successful product search, yet measuring relevance of items to queries is one of the most challenging tasks in product information retrieval, and quality of product search is highly influenced by the precision and scale of available relevance-labelled data. In this paper, we present an array of techniques for leveraging Large Language Models (LLMs) for automating the relevance judgment of query-item pairs (QIPs) at scale. Using a unique dataset of multi-million QIPs, annotated by human evaluators, we test and optimize hyper parameters for finetuning billion-parameter LLMs with and without Low Rank Adaption (LoRA), as well as various modes of item attribute concatenation and prompting in LLM finetuning, and consider trade offs in item attribute inclusion for quality of relevance predictions. We demonstrate considerable improvement over baselines of prior generations of LLMs, as well as off-the-shelf models, towards relevance annotations on par with the human relevance evaluators. Our findings have immediate implications for the growing field of relevance judgment automation in product search.
\end{abstract}



\maketitle

\section{Introduction}
\label{sec:intro}


Can LLMs replace humans as evaluators of query-item relevance? LLMs can act as tools for automated assistance to human evaluation, as well as candidates for completely replacing human involvement \cite{faggioli23}. The introduction of increasingly larger and more sophisticated language models has made machine relevance evaluation even closer to the human performance.

In this paper we examine methods for improving the performance of LLMs towards parity with human performance. Using a unique dataset of e-commerce data we examine and report on an array of techniques for finetuning LLMs for relevance judgments in product search. The finetuned models can be used to scale relevance labelling at scale, with accuracies approaching that of the gold standard labels, defined as the judgments produced by an informed user. 

Our results in section (\ref{sec:methods}) show that Low Rank Adaptation (LoRA) \cite{hu2022lora} of intermediate-sized with the training datasets sized in the order of $10^6$ perform better than full model finetuning. This observation emphasizes the tradeoff between the model and data size. For smaller training datasets, LoRA is expected to align with data more efficiently than full finetuning. We also find that including additional item textual data in the form of description, as expected, improves performance in larger multi-billion parameter models, but not in smaller ones of size in the order of hundred million parameters. 

Our training dataset is comprised of $6,055,251$ QIPs, $708,895$ queries and $2,664,741$ items. The dataset, similar to those in the public domain (e.g. \cite{reddy2022shopping}), is imbalanced towards the most relevant label, but unlike the ESCI dataset introduced in \cite{reddy2022shopping}) has an almost complete coverage of longer item description information.\footnote{The dataset introduced in \cite{reddy2022shopping} has an ``item\_description'' coverage close to 48\%.} Our relevance models are classifiers that are trained on inputs of concatenated item textual features and query along with relevance labels. On a three-class graduation of relevance labels (relevant, related, irrelevant) we achieve micro f1 scores of 88.8\% and 89.6\% with two larger language models. There are improvements up to 4.6\% in f1 micro scores compared to the performance of pretrained off-the-shelf decoder-only LLMs (see tables  \ref{tab:res_llama}, \ref{tab:res_mistral}, and \ref{tab:res_bert}).

We show that the balance between number of queries per item and items per query in the relevance labelled training dataset change the extent to which the item textual data is included in training, and hence both the number of queries available for each item included and more scrutinized item per query parallel, are correlated with the training performance. With fewer queries per item included in the dataset, adding item textual information is less helpful. On the other hand, when each item is paired with multiple queries in the training dataset, the performance improves. The query segment of the QIP input data is often much shorter than the item textual attributes and enriching and augmenting the query part of input to the model is expected to improve the performance.  

We also demonstrate the regions of LoRA rank and $\alpha$, as well as LoRA dropouts, where two 7B LLMs perform best, showing that, unlike conventional wisdom, with the size of the training dataset and LLMs in question, $\alpha$s in the range of 1 to 2 times LoRA rank perform better than alternatives. LoRA finetuning, in turn, results in higher f1 scores than full model finetuning in our trainings. 

In our finetuning experiments we have leveraged an array of decoder-only larger language models, which are increasingly used for classification and generative purposes in lieu of encoder-decoder models such as the FLAN-T5 class \cite{flan22} used in \cite{kang23llm}.

Finally, to further test the performance of our models in section~(\ref{sec:rel}) we re-enact a set of 718 feature launch experiments based on human ranking relevance judgments on sample QIPs with item rankings for variation and control per query. Decisions are made based on nDCG comparisons at 1,5 and 10, between control and variation QIPs. We show that the usage of LLMs as evaluators of QIP relevance matches human evaluators to the point that no feature launch judged by human evaluators is reversed by our finetuned LLMs. Relevance models trained on the aforementioned dataset achieve agreement with human evaluators' nDCG\@1-based comparison of control and variation, in up to 89\% of the 718 feature launch experiments. 

Such levels of agreement between human evaluation judgment and automated relevance evaluation suggest LLMs can be reliably trained to act as cost effective and fast replacements for human relevance judgments. For fully realizing the potential of LLMs for relevance judgment, a further host of training techniques, such as self-distillation and effective pretraining on product data, can be applied; we list them in conclusion as future work. 


\section{Related Work}
\label{sec:related}

With the recent introduction of increasingly sophisticated large language models \cite{transformer,mistral23,llama23}, the training of cross-encoder models\cite{bert} of relevance judgment on human-labelled datasets \cite{reddy2022shopping} has become competitive with the human performance \cite{thomas2023}. 

Traditionally, progress in information retrieval (IR)~\cite{nir,ir} is driven by extensive data collection efforts by human experts for evaluating the quality of search and retrieval results. The difficulties of setting up such an evaluation procedure are well studied~\cite{faggioli23}. Such a process necessitates selecting items (documents), queries and a group of human experts to judge the QIPs. This can be extremely expensive and time consuming. Human evaluators' performance is difficult to calibrate and standardize. In fact, it is not certain if human evaluators assigned can always outperform language models \cite{faggioli23}. Hence, large language models present a unique opportunity for automating the process at scale. 

The existing work on the application of LLMs in relevance evaluation constitutes of prior art in at least three subfields: crowd sourcing methods, query-product classification techniques used for automation and the novel realm of LLM usage for producing automated relevance evaluations on par with the gold standard. 

\subsection{Crowd Sourcing}
Crowd sourcing is a traditional solution for relevance labelling~\cite{alonso2009can} which has become a more practical procedure using online tools for orchestration. Relegating relevance judgment to human evaluators has been an effective tool for relevance evaluation of IR systems~\cite{maddalena16}. Crowd sourcing, being cheaper than expert human evaluators, faces numerous challenges: quality and reliability of workers can be hard to check and the particular setup of the task, for example the length of the judgment and the extent of item data shown to evaluators, can create undesired biases in the label data. It has been shown that with careful creation of evaluation tasks and with effective guardrails for quality, it is possible to collect relevance data at scale~\cite{alonso2009can,blanco11,alonso2011design,gadiraju2019crowd,kazai2013analysis,maddalena16,nouri2020mining,sheshadri2013square,tamine2017impact}. Nevertheless, in search and retrieval feature evaluation and relevance testing, crowd sourcing can still be prohibitively expensive, furthermore, the process of human labelling is difficult to standardize. Crowd labelers' perception of query-item relevance can rely on a complex combination of subjective perception and objective data that differs from one evaluator to the next.  


\subsection{Query-Product Classification}
Relevance classifiers of query-item pairs (QIPs)\cite{jiang19} can help to automate the process of relevance judgment. To reliably train such classifiers, high quality relevance-labelled QIPs are necessary \cite{reddy2022shopping}. Apart from data, language model embeddings \cite{li2021embedding}\cite{liu2021que2search} and LLMs' next token prediction capabilities with cross-encoder style \cite{shan23} input, and a classifier layer added to the LLM architecture, can be leveraged to construct an automated pipeline akin to human evaluators. We examine a range of the aforementioned possibilities in the following sections. 

In the process of training, crowd labels replace oft-used engagement data \cite{DSSM,yao2021learning,yao2021learning} for training search and retrieval relevance estimation and ranking. The issues with malleability and standardization of crowd evaluation \cite{faggioli23} we enumerated above can be resolved in the controlled settings of LLM classification. Yet, the need for a starting seed dataset of high quality labeled QIPs remains. 

\subsection{LLM Use in Product Search}

With the introduction of large language models, a revolution in semantic search and retrieval has occurred. LLMs are prime candidates for relevance classifiers: they are capable of teasing out long-memory dependencies and contextual relations. LLMs are shown to be reliable predictors of customer preference \cite{kang23llm,bertimp}. Earlier versions of encoder-decoder models are used in two-tower embedding-based semantic retrieval designs \cite{zhang2020towards} that are currently deployed in production systems for retrieval \cite{magnani2022semantic,muhamed2023web, nigam2019semantic,he2023que2engage,jha2023unified,li2021embedding,zhang2020towards}.


In comparison to the two-tower dual encoder architecture, the cross-encoder architecture~\cite{lu2022ernie} is shown to achieve higher performance, and cross encoder models therefore are used in the ranking phase of IR systems~\cite{nogueira2019passage,yang2019simple,qu2021rocketqa,muhamed2023web}. 

To help overcome challenges in the introduction of LLMs into product search \cite{campos2023overview}, designs based on the cross-encoder architecture \cite{lu2022ernie} have been introduced where one single LLM takes query and a subset of item (and query) attributes, mainly as textual features, as input and a classification layer assembled on top of the encoder-only LLM predicts the classification labels, this is the architecture we employ in our designs. Techniques such as self-distillation \cite{xuy20} are applied when the size of training dataset is limited by the human labelling capacities and scale, and the latency and cost requirements of LLMs necessitate their usage offline as a teacher model for self-distillation.





\section{Methodology}
\label{sec:methods}

In the following, we outline the data variations and model varieties we have employed for relevance evaluation training. 

\subsection{Data}
\label{sec:data}


Throughout several years of developing new capabilities for improving the search experience on our commercial search engine, we collected millions of relevance annotations for point-wise query item pairs (QIPs) both internally and through external vendors. A QIP is presented to an annotator as a free-text query, an image of the item, its title and a few attributes, such as brand and size. Annotators are also provided a link to the item page that can be accessed in case of doubt. The annotations are tagged according to the following 3-point scale: \[
2 = \text{Relevant}, 1 = \text{Related}, 0 = \text{Irrelevant}.\] Table~\ref{tab:datum} shows an example for each relevance level.

From this collection of QIPs, we extracted, at random, two sets of sizes 6,055,251 and 66,072, which are used for training and testing our models respectively.

Examples of the three label categories are included in Table (\ref{tab:datum}). The total number of queries and items in the dataset and the distribution of labels among the QIPs are included in tables (\ref{tab:query_item}) and (\ref{tab:num_labels}). 

\begin{table*}
    \centering
    \begin{tabular}{ll}
        \hline
        Query & graco 5 in 1 crib mattress \\
        \hline
        Relevant item & Graco Benton 5-in-1 Convertible Baby Crib with Drawer, White \\
        Related item & Graco 5" Crib and Toddler Mattress with Ultra-Soft, Water-Resistant, Removable Outer Cover \\
        Irrelevant item & Graco Pack 'n Play Playard Newborn Fitted Sheets, 2 Pack, Gust\\
        \hline
    \end{tabular}
    \vspace{.05in}\caption{Example of a query with three different labels of candidate items.}
    \label{tab:datum}
\end{table*}

\begin{table}
    \centering
    \begin{tabular}{lr}
    \hline
       total \# of queries  & 708,895\\
       total \# of items    & 2,664,741\\ 
       \hline
    \end{tabular}
    \vspace{.05in}\caption{Total number of queries and items in training data}
    \label{tab:query_item}
\end{table}

\begin{table}
    \centering
    \begin{tabular}{lrr}
    \hline
      label & \multicolumn{1}{c}{\#} & \multicolumn{1}{c}{\%}  \\ \hline
      Relevant     & 3,887,072& 64.19\%\\
      Related      & 1,191,913& 19.68\% \\
      Irrelevant   &   976,266& 16.12\% \\
      \hline
    \end{tabular}
    \vspace{.05in}\caption{Distribution of dataset labels, total 6,055,251 }
    \label{tab:num_labels}
\end{table}

We faced the important question of the accuracy of human labels. To account for imperfections of the human judgment process, the labels of the test set were collected for a second time. If these secondary labels disagreed with the first set of labels, a third set of human labels were produced to decide on the final label.

Throughout the experiments conducted in this paper, we consider these majority-vote labels to be the golden labels, against which both the models and individual human labelers (from the first two rounds of annotations) are evaluated. 

Human performance measures are included in Table (\ref{tab:human}). Later, we use the same measures to demonstrate the gap between our finetuned LLMs and the human performance.

\begin{table}
    \centering
    \begin{tabular}{l|cccc}
         & f1 class-0 & f1 class-1 & f1 class-2 & f1 micro \\ \hline
                Human perf. & 0.890 & 0.862 & 0.968 & 0.937 \\ \hline
        
    \end{tabular}
    \vspace{.05in}\caption{Human Evaluators Performance, Initial labels versus Majority (of Three Human-produced Labels)}
    \label{tab:human}
\end{table}

\begin{figure}
    \centering
    \includegraphics[width=.97\linewidth]{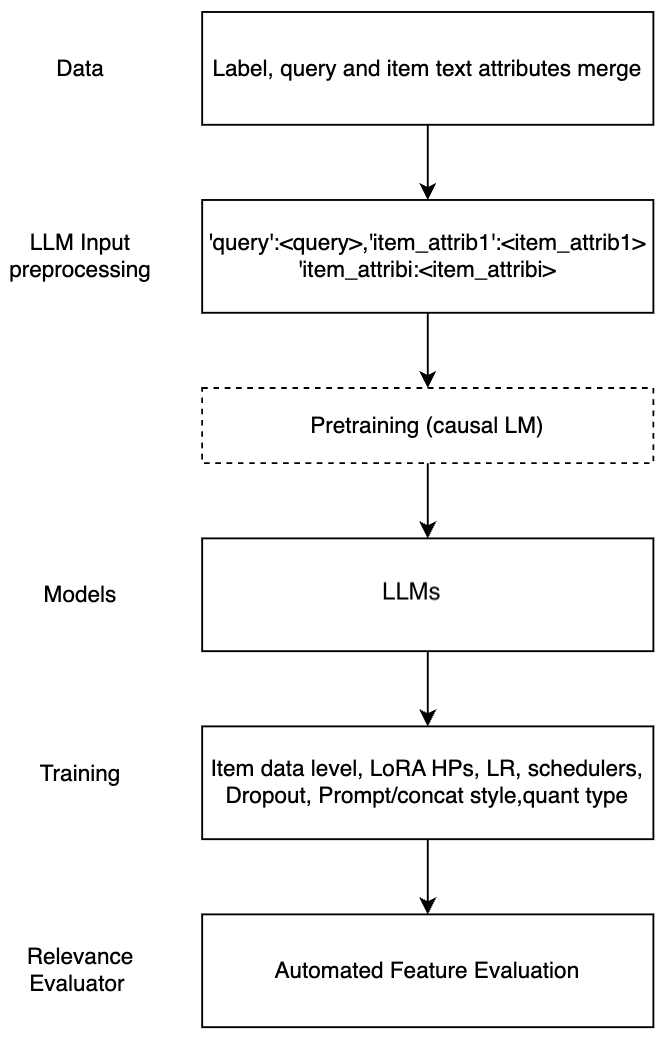}
    \caption{LLM Relevance classification training pipeline}
    \label{fig:llm_relevance}
\end{figure}


This study uses three variations of item data in relevance training, details of which are provided in Table~\ref{tab:item_data}. We investigate the impact of including additional textual data on the accuracy of relevance prediction by training models with these QIPs settings. We have trained models with the three variations of item data included in Table~\ref{tab:item_data}.

This approach allows us to systematically analyze the influence of both information richness and query distribution on relevance prediction accuracy. We have used three variations of item data in relevance training, a summary of which is included in Table~\ref{tab:item_data}. To examine the effect of including additional textual data on the accuracy of relevance prediction, we have trained models with QIPs where either \#1) only item title, product type, brand, color and gender or \#2) those item attributes in \#1 plus a longer textual description of the item. To further examine the interplay between the distribution of query per item and training results we have tried two variations of the item description addition: one in which each item with description is connected\footnote{i.e. the QIP is present in the dataset.} to more queries (approximately three on average in the training dataset) and another variation in which the same rate is lower at approximately 1 query per item in the training dataset. Hence, we explore the interplay between the query-per-item distribution and training results through two variations of the enriched description addition.

It is necessary to note that the variations in item attribute data available to human evaluators are key to the result of judgments, similarly we expect similar variations to impose a range of relevance prediction accuracies. One of the main advantages of LLM usage for relevance judgment on scale is that more sophisticated models are expected to be capable of exploiting longer item descriptions that are not consumable by the human evaluator in a short period of time. 

To take stock, we have experimented with the inclusion of various levels of item attribute data \cite{sun23} in training, as well as query perturbations \cite{chaudhary23} for making the training more robust. We expected that adding noise to the queries in the training process, akin to adversarial perturbations that can help making training more robust, particularly because the item attribute data is more diverse and voluminous compared to that of the queries. The query's textual context is considerably smaller than the item textual descriptions, therefore having a QIP set where more queries are linked with a given item is expected to improve performance.\footnote{Adding query attributes to the query itself is another venue for enriching the training dataset, this is relegated to the future work.}

\begin{table}
    \centering
    \begin{tabular}{cl}
    \hline
       \#1  & {\it title}, {\it product\_type}, {\it brand}, {\it color}, {\it gender} \\ \hline
        \#2 & \#1 + {\it item description}, query per item w descr $\approx$ 3\\ \hline
        \#3 & \#1 + {\it item description}, query per item w descr $\approx$  1 \\ \hline
        \hline
    \end{tabular}
    \vspace{.05in}\caption{Item Data Variations in Training}
    \label{tab:item_data}
\end{table}

\subsection{Relevance Models}
\label{sec:models}

We have employed 7B versions of two open-source Large Language Modles henceforth called LLM3 \cite{llama23} and LLM4 \cite{mistral23} in our training along with bidirectional encoder transformer model 1  \cite{bert},\cite{bertimp}, \cite{reimers19}, henceforth called LLM1, and bidirectional encoder transformer model 2 \cite{he2023debertav} henceforth called LLM1, as baselines for larger language models. 

We have trained these models using PyTorch\cite{NEURIPS2019_9015}, PyTorch Lightning \cite{Falcon_PyTorch_Lightning_2019} and, Huggingface Transformers' framework \cite{wolf2020huggingfaces}. 

We have adopted midsized LLMs, because, as we show later, given the size of the training dataset used, performance results start to decline with increasing the rank of LoRA, in fact, the models fare better with Low Rank adaptation compared to full model finetuning. For the very same reason, we have not included the full set of performance results from models with their number of parameters larger than 7B.

The dichotomies applied in this study, include:

\begin{enumerate}
\item LoRA (using library {\tt peft} \cite{peft}) versus full model finetuning
\item Prompting versus last token-based classification
\item Item verbosity levels in training
\item Hyper parameter choices such as
\begin{enumerate}
\item (LoRA) Dropout \cite{dropout}
\item LoRA rank and alpha, layers trained on
\item Learning rate (lr) and the lr scheduler 
\item Training optimizer \cite{kingma2014adam} variations 
\end{enumerate}
\item Model choice \& model quantization
\item Label class weights
\end{enumerate}

In the following, we present the model training procedure for a number of LLMs discussed. 

\subsubsection{LLM1}
Previous studies have shown that the cross encoder based LLM1 model \cite{luan2021sparse} performs well on solving semantic retrieval problems. To address the specific challenge of e-commerce product relevance search, we have employed a cross encoder model in our design. We experimented with finetuning our model using two variations of the LLM1 model as the starting checkpoints: 1) the LLM1-Base uncased model from Huggingface, and 2) this same model further pretrained on the Walmart Search corpus. 

{\it Walmart Pretrained \iffalse BERT\fi LLM1:} Since the LLM1-base model was trained on general domain text data, we aimed to enhance its performance on e-commerce related texts by conducting additional language model training using Walmart product catalog. We chose the masked language modeling task with 15\% masking rate, also included a binarized orders prediction task. We interleaved the training between these two tasks, with each batch consisting of only one task. We utilized one year of Walmart search behavioral data for the binarized orders prediction training.

{\it Cross Encoder Architecture:} We finetune our cross-encoder model using human-labeled query-item relevance pairs and aim to classify the relevance of these pairs. As shown in Figure~(\ref{fig:llm_relevance}), we begin by merging all query-item features into a single concatenated string. This string is then fed into the LLM1 encoder. After the LLM1 encoder pass, we apply dropout and then non-linear layers to transform the embedding into a three-dimensional classification vector. Lastly, we apply a Softmax layer to convert this vector into a three-class relevance probability vector, which is optimized using the cross-entropy loss. 

\subsubsection{\iffalse DeBERTa\fi LLM2}
LLM2, an improvement upon LLM1, is used, similar to the LLM1 implementation described above. 

\subsubsection{\iffalse Llama2\fi LLM3}
We finetuned LLM3-7B as a part of cross-encoder architecture for predicting relevance between query and items. We experimented with finetuning a classifier (a dense classification layer superimposed on LLM3 7B), with and without the LoRA implementation. With Low Rank Adaptation of LLM3 we explored a range of rank and $\alpha$s that dictate the size of linear approximation to the model, and the ratio to which the linear differential approximations were included in training \cite{hu2022lora}. The classification itself is based on the ``last token'' that is not a padding token. We ran experiments with increasing the size of model parameters used in training, based on LoRA rank, to find the optimal model parameter size for classification based on number of labels and training data points. 

\subsubsection{\iffalse Mistral\fi LLM4}
Similar to LLM1 and LLM3, we tried full fine tuning as well as LoRA fine tuning of LLM4-7B-v0.1 using various hyper parameters and learning rate schedulers. We experimented on the following aspects: various learning rate schedulers, including and excluding product description during training and testing phases, product feature dropouts, introducing random negatives, domain adaptation pretraining of LLM4-7B as causal LM on walmart data followed by full fine tuning, and domain adaptation pretraining of LoRA adapters on LLM4-7B as causal LM on Walmart data, followed by LoRA fine tuning of the domain adapted model.

\section{Results: Training Findings, Discussion}
\label{sec:models_train}

In the following we outline the findings from relevance training experiments in Tables \ref{tab:res_bert}, \ref{tab:res_llama} and \ref{tab:res_mistral}. We compare the outcomes contingent on model choice, training data level, LoRA hyper parameters, learning rates, learning rate scheduler type, attention dropout, pretraining being present or not, as well as class weights being applied.\footnote{For the variables with continuous support and numerous manifestations, such as dropout and LoRA $\alpha$ and rank we have included the best results, excluding the sub-optimal values that were not useful for comparison pruposes.}

\subsection{\iffalse BERT\fi LLM1}
Table~\ref{tab:res_bert} contains the results for LLM1 and LLM2. Models B1., B2. and B3. in Table \ref{tab:res_bert} differ only in drop out rates from $0.1$ to $0.05$, with little change in outcome. Model B4. is trained starting from a Walmart pretrained LLM1, with no class weights (unlike models B1. to B3.) and uses variation \# 1 in table \ref{tab:item_data} with no verbose item description used in training. Models B5. and B6., however, do use verbose description (option \#3 in Table \ref{tab:res_bert} ) and as a result show a boost in performance compared to models B3. and B4. Finally models B7. to B11. use public version of LLM1, for finetuning, instead of pretrained ones, and show degradation in performance compared to models B1. to B6. 

LLM2 experiments D1. and D2. present LoRA trainings with rank=64, $\alpha$=16, on all linear modules and a learning rate of $1e-4$ and learning rate scheduler with linear warmup. The comparison shows that including item descriptions has improved the model's performance. 

\subsection{\iffalse Llama\fi LLM3}
Table~\ref{tab:res_llama} presents the results of experiments involving LLM3 7B. We examined LoRA ranks in the range of 16 to 512, noting that increasing LoRA rank beyond the range of 256 did not improve the model performance. For rank =256, LoRA $\alpha$s closer to half of rank yielded the highest f1 scores. Similar experiments, comparing LoRA dropouts 0.1 and 0.05, found the latter to be preferable. The learning rate for all the LLM3 experiments is $1e-4$, and decreasing the learning rate to $1e-5$ did not improve the performance. Compared to the experiment L.3, decreasing $\alpha$ further to 64, worsened, and including class weights, improved the results. Removing added noise to the queries degraded the performance, as well as using item data without description (item data \#1 in Table \ref{tab:item_data}). Comparing L4. to L3., we find that resolving the disagreement between two sets of human evaluators via a third vote significantly improves the results on the reformed test set, implying the potential for labelling improvements over the existing human judgment labels.\footnote{The procedure for improving the test set label accuracy is described in section \ref{sec:data}.}

Finally, LoRA training shows better performance compared to full finetuning of LLM3 7B with our dataset (L11. versus L4.)

\subsection{\iffalse Mistral\fi LLM4}
Results are included in Table~\ref{tab:res_mistral}. Including item description during the training helped improve the test metrics, furthermore a comparison between M1. and M2. shows that training on the dataset with higher number of queries per item yields better performance. A similar comparison between LLM4 experiments M3.1 and M3.2 shows training on item attribute dataset \#2 in Table \ref{tab:item_data} improves over dataset \#3. 


We experimented with random feature dropouts such as dropping the product\_type, product\_description and brand along with random negatives at rates 0.01 and 0.3. Feature dropouts improved f1 metrics and reduced the training time, since the total sequence length consumed by the model was reduced. The reduction in training time mainly came from product description dropouts. Descriptions are lengthier features compared to product title, product\_type or brand.


The Domain adaptation pretraining of LoRA adapters, followed by LoRA finetuning did not show a significant improvement over the standalone LoRA finetuning. Also Similar to LLM3, LoRA trainings on LLM4 result in better performance compared to full model finetuning. 

The Reduce on plateau learning scheduler showed the best results when we ran validation step after training on every 10\% of the data and adjusted the learning rate. When validation micro-f1 score did not improve by $0.001$, we reduced the learning rate by a factor of 2. While pretraining LoRA adapters, we used a cosine annealing rate scheduler with warm restarts after every 115k steps, where each of the steps computed gradients over 256 sentences from an e-commerce catalog and query log.

\vspace{.1 in}
For models in tables \ref{tab:res_llama} and \ref{tab:res_mistral} AdamW optimizer is used for training, also for both classes last token is used for classification. Models M1 and M3.x in table \ref{tab:res_mistral} use reduce\_on\_plateau learning rate schedulers, while model M2. uses a linear warmup + cosine scheduler. In Table \ref{tab:res_mistral}, learning rates are $1e-5$.

\begin{table*}
    \centering
    \begin{tabular}{l|cccc}
        Methods & f1 class-0 & f1 class-1 & f1 class-2 & f1 micro \\ \hline
         baseline, LLM1-base Cross-encoder   & 0.737	& 0.695 & 0.913	& 0.843 \\ \hline
         - L0. baseline, LLM3 7B pretrained   & 0.725	& 0.662 & 0.923	&  0.853 \\ \hline
       + L1. LoRA rank = 256, $\alpha$ =128, all linear layers, item data \#2  &   0.763 & 0.716 & 0.928 & 0.862 \\ \hline
       + L2. LoRA rank = 256, $\alpha$ =128, all linear layers, item data \#2, lr=1e-5 &  0.754 & 0.705 & 0.925 & 0.862 \\ \hline
       + L3. LoRA rank = 256, $\alpha$ =128, all linear layers, item data \#2, lora\_dropout=0.05 &   0.765 & 0.716 & 0.929 & 0.868 \\ \hline
       + L4. LoRA rank = 256, $\alpha$ =128, all linear layers, data \# 2, lora\_dropout=0.05, test maj. vote &   {\bf 0.791} & 
 {\bf 0.758} & {\bf 0.941} & {\bf 0.888}	
 \\ \hline
       + L5. LoRA rank = 256, $\alpha$ =64, all linear layers, item data \#2, lora\_dropout=0.05 &   0.763 & 0.712 & 0.929 & 0.867 \\ \hline
       + L6. L3. with weighted classes (1:0.9:0.5) &   0.767 & 0.715 & 0.928 & 0.865 \\ \hline
       + L7. L3. without query noise &   0.762 & 0.714 & 0.928 & 0.866 \\ \hline
       + L8. same as L3. but training without description, item data \# 1 &  0.761 & 0.709 & 0.927 & 0.863\\ \hline
       + L9. training and test wout description, item data \#1 , test maj. vote &  0.782 & 0.745 & 0.938 & 0.883 \\ \hline
       + L10. L3. with LoRA rank = 512 &   0.751 & 0.709 & 0.927 & 0.863\\ \hline
       + L11 LLM3-7B Full finetune (attention\_dropout=0.4) & 0.786 & 0.758 & 0.944 & 0.889 \\ \hline
       \# Human evaluator performance & 0.890 & 0.862 & 0.968 & 0.937 \\ \hline
       
    \end{tabular}
    \vspace{.05in}\caption{Ablation Studies for LLM3\iffalse Llama2\fi}
    \label{tab:res_llama}
\end{table*}

\begin{table*}
    \centering
    \begin{tabular}{l|ccccc}
        Methods & f1 class-0 & f1 class-1 & f1 class-2 & f1 micro \\ \hline
        baseline, LLM1-base Cross-encoder   
        & 0.737	& 0.695 & 0.913	& 0.843 \\ \hline
        - M0. baseline, LLM4-7B pretrained
        & 0.669	& 0.679 & 0.924	& 0.853 \\ \hline
       + M1. LLM4-7B-v0.1 (attention\_dropout=0.4) , \#2 item descr\ dataset 
       & 0.789	& 0.766 &	0.944 & 0.891 \\ \hline
       + M2. LLM4-7B-v0.1 (attention\_dropout=0.4), \#3 item descr dataset   
       & 0.787 & 0.752 & 0.941 & 0.886 \\ \hline
       + M3. LLM4-7B-v0.1 LoRA (r=256, $\alpha$=128, d=0.05, all linear layers), \#3 item  dataset  
       & 0.767 & 0.739 & 0.937 & 0.879 \\ \hline
       + M3.1 LLM4-7B-v0.1 LoRA (M3.+ $\alpha$=256+ embedding layer), \#2 item  dataset    
       & 0.800 & 0.767 & 0.947 & 0.895 \\ \hline
       + M3.2. LLM4-7B-v0.1 LoRA (same as M3.), \#2 item  dataset
       & 0.799 & 0.771 & 0.947 & 0.896 \\ \hline
       + M3.3. LLM4-7B-v0.1 LoRA (same as M3.), \#3 item  dataset
       & 0.795 & 0.768 & 0.946 & 0.894 \\ \hline
       + M3.4. LLM4-7B-v0.1 LoRA (same as M3.), \#2 item  dataset, w feature dropout+RN=0.01
       & {\bf 0.804} & {\bf 0.770} & {\bf 0.947} & {\bf 0.896} \\ \hline
       + M3.4.1. LLM4-7B-v0.1 LoRA (same as M3.), \#1 item  dataset, w feature dropout+RN=0.01
       & 0.798 & 0.767 & 0.945 & 0.894 \\ \hline
       + M3.5. LLM4-7B-v0.1 LoRA (same as M3.),\#2 item  dataset, w feature dropout+RN=0.3   & 0.798 & 0.771 & 0.947	& 0.894 \\ \hline

       + M3.6. LLM4-7B-v0.1 LoRA (same as M3.), Pretrain 20\% + FT prams same as 3.5. & 0.786 & 0.758 & 0.944 & 0.889 \\ \hline
        \# Human evaluator performance & 0.890 & 0.862 & 0.968 & 0.937 \\ \hline
       
    \end{tabular}
    \vspace{.05in}\caption{Ablation Studies for \iffalse Mistral\fi LLM4}
    \label{tab:res_mistral}
\end{table*}

\begin{table*}
    \centering
    \begin{tabular}{l|cccc}
        Methods & f1 class-0 & f1 class-1 & f1 class-2 & f1 micro \\ \hline
        baseline, LLM1-base Cross-encoder   & 0.737	& 0.695 & 0.913	& 0.843 \\ \hline
       + B1. Walmart Pretrained LLM1-base with class-weights, dropout=0.1  & 0.763 & 0.718 & 0.921 & 0.858  \\ \hline
       + B2. Walmart Pretrained LLM1-base with class-weights, dropout=0.2 & 0.762 & 0.719 & 0.922 & 0.858  \\ \hline
       + B3. Walmart Pretrained LLM1-base with class-weights, dropout=0.05 & 0.763 & 0.719 & 0.921 & 0.858  \\ \hline
       +B4. Walmart Pretrained LLM1-base no class-weights, dropout=0.1 &  0.763 &  0.724 & 0.932 & 0.873 \\ \hline
       +B5. Walmart Pretrained LLM1-base with class-weights, dropout=0.1, w. item description \#2 & 0.761 & 0.723 & 0.925 & 0.861  \\ \hline
       + B6. Walmart Pretrained LLM1-base no class-weights, dropout=0.1, w. item description \#2 & {\bf 0.765} & {\bf 0.720} & {\bf 0.933} & {\bf 0.873}  \\ \hline
       + B7. LLM1-base no class-weights, dropout=0.1 & 0.720 & 0.692 & 0.922 & 0.855  \\ \hline
       + B8. LLM1-base with class-weights, dropout=0.1& 0.714 & 0.691 & 0.922 & 0.853  \\ \hline
       + B9. LLM1-base no class-weights, dropout=0.1 w. item description \#3 & 0.727 & 0.706 & 0.927 & 0.861  \\ \hline
       + B10. LLM1-base with class-weights, dropout=0.1 w. item description \#3 & 0.724 & 0.670 & 0.905 & 0.831  \\ \hline
       +11. LLM1-base no class-weights, dropout=0.1 w. item description \#2 & 0.725 & 0.704 & 0.927 & 0.860  \\ \hline \hline
       + D1. LLM2 training w no item description, LoRA & 0.774 & 0.7380 & 0.935 & 0.878  \\ \hline 
       + D2. LLM2 training w item description \# 2, LoRA & {\bf 0.782} & {\bf 0.753} & {\bf 0.940} & {\bf 0.885}  \\ \hline \hline
        \# Human evaluator performance & 0.890 & 0.862 & 0.968 & 0.937 \\ \hline
    \end{tabular}
    \vspace{.05in}\caption{Ablation Studies for \iffalse BERT/DeBERTa\fi LLM1/LLM2, used as Baseline for Larger Models}
    \label{tab:res_bert}
\end{table*}


\begin{table*}[!htbp]
\centering
\begin{tabular}{|cc|rrrc|rrrc|rrrc|}
\hline
\multicolumn{2}{|c|}{\multirow{3}{*}{\textbf{}}}                         & \multicolumn{4}{c|}{\textbf{nDCG@1}}                                                                                                                                                     & \multicolumn{4}{c|}{\textbf{nDCG@5}}                                                                                                                                                     & \multicolumn{4}{c|}{\textbf{nDCG@10}}                                                                                                                                                    \\ \cline{3-14} 
\multicolumn{2}{|c|}{}                                                   & \multicolumn{3}{c|}{\textbf{Human}}                                                                 & \multirow{2}{*}{\textbf{\begin{tabular}[c]{@{}c@{}}Combined\\ Score\end{tabular}}} & \multicolumn{3}{c|}{\textbf{Human}}                                                                 & \multirow{2}{*}{\textbf{\begin{tabular}[c]{@{}c@{}}Combined\\ Score\end{tabular}}} & \multicolumn{3}{c|}{\textbf{Human}}                                                                 & \multirow{2}{*}{\textbf{\begin{tabular}[c]{@{}c@{}}Combined\\ Score\end{tabular}}} \\ \cline{3-5} \cline{7-9} \cline{11-13}
\multicolumn{2}{|c|}{}                                                   & \multicolumn{1}{c|}{\textbf{+}} & \multicolumn{1}{c|}{\textbf{=}} & \multicolumn{1}{c|}{\textbf{-}} &                                                                                    & \multicolumn{1}{c|}{\textbf{+}} & \multicolumn{1}{c|}{\textbf{=}} & \multicolumn{1}{c|}{\textbf{-}} &                                                                                    & \multicolumn{1}{c|}{\textbf{+}} & \multicolumn{1}{c|}{\textbf{=}} & \multicolumn{1}{c|}{\textbf{-}} &                                                                                    \\ \hline
\multicolumn{1}{|c|}{\multirow{3}{*}{\textbf{\iffalse BERT\fi LLM1}}}        & \textbf{+} & \multicolumn{1}{r|}{61}         & \multicolumn{1}{r|}{66}         & \multicolumn{1}{r|}{0}          & \multirow{3}{*}{0.840}                                                              & \multicolumn{1}{r|}{129}        & \multicolumn{1}{r|}{50}         & \multicolumn{1}{r|}{0}          & \multirow{3}{*}{0.852}                                                              & \multicolumn{1}{r|}{159}        & \multicolumn{1}{r|}{68}         & \multicolumn{1}{r|}{0}          & \multirow{3}{*}{0.819}                                                              \\ \cline{2-5} \cline{7-9} \cline{11-13}
\multicolumn{1}{|c|}{}                                      & \textbf{=} & \multicolumn{1}{r|}{13}         & \multicolumn{1}{r|}{444}        & \multicolumn{1}{r|}{20}         &                                                                                    & \multicolumn{1}{r|}{10}         & \multicolumn{1}{r|}{355}        & \multicolumn{1}{r|}{30}         &                                                                                    & \multicolumn{1}{r|}{12}         & \multicolumn{1}{r|}{282}        & \multicolumn{1}{r|}{32}         &                                                                                    \\ \cline{2-5} \cline{7-9} \cline{11-13}
\multicolumn{1}{|c|}{}                                      & \textbf{-} & \multicolumn{1}{r|}{0}          & \multicolumn{1}{r|}{16}         & \multicolumn{1}{r|}{98}         &                                                                                    & \multicolumn{1}{r|}{0}          & \multicolumn{1}{r|}{16}         & \multicolumn{1}{r|}{128}        &                                                                                    & \multicolumn{1}{r|}{0}          & \multicolumn{1}{r|}{18}         & \multicolumn{1}{r|}{147}        &                                                                                    \\ \hline
\multicolumn{1}{|c|}{\multirow{3}{*}{\textbf{\iffalse DeBERTa\fi LLM2}}}     & \textbf{+} & \multicolumn{1}{r|}{61}         & \multicolumn{1}{r|}{54}         & \multicolumn{1}{r|}{0}          & \multirow{3}{*}{0.875}                                                              & \multicolumn{1}{r|}{127}        & \multicolumn{1}{r|}{57}         & \multicolumn{1}{r|}{0}          & \multirow{3}{*}{0.847}                                                              & \multicolumn{1}{r|}{158}        & \multicolumn{1}{r|}{65}         & \multicolumn{1}{r|}{0}          & \multirow{3}{*}{0.822}                                                              \\ \cline{2-5} \cline{7-9} \cline{11-13}
\multicolumn{1}{|c|}{}                                      & \textbf{=} & \multicolumn{1}{r|}{13}         & \multicolumn{1}{r|}{462}        & \multicolumn{1}{r|}{13}         &                                                                                    & \multicolumn{1}{r|}{12}         & \multicolumn{1}{r|}{351}        & \multicolumn{1}{r|}{28}         &                                                                                    & \multicolumn{1}{r|}{13}         & \multicolumn{1}{r|}{283}        & \multicolumn{1}{r|}{30}         &                                                                                    \\ \cline{2-5} \cline{7-9} \cline{11-13}
\multicolumn{1}{|c|}{}                                      & \textbf{-} & \multicolumn{1}{r|}{0}          & \multicolumn{1}{r|}{10}         & \multicolumn{1}{r|}{105}        &                                                                                    & \multicolumn{1}{r|}{0}          & \multicolumn{1}{r|}{13}         & \multicolumn{1}{r|}{130}        &                                                                                    & \multicolumn{1}{r|}{0}          & \multicolumn{1}{r|}{20}         & \multicolumn{1}{r|}{149}        &                                                                                    \\ \hline
\multicolumn{1}{|c|}{\multirow{3}{*}{\textbf{\iffalse Llama\fi LLM3 L.4}}}   & \textbf{+} & \multicolumn{1}{r|}{65}         & \multicolumn{1}{r|}{51}         & \multicolumn{1}{r|}{0}          & \multirow{3}{*}{\bf 0.887}                                                              & \multicolumn{1}{r|}{129}        & \multicolumn{1}{r|}{59}         & \multicolumn{1}{r|}{0}          & \multirow{3}{*}{0.847}                                                              & \multicolumn{1}{r|}{160}        & \multicolumn{1}{r|}{52}         & \multicolumn{1}{r|}{0}          & \multirow{3}{*}{0.851}                                                              \\ \cline{2-5} \cline{7-9} \cline{11-13}
\multicolumn{1}{|c|}{}                                      & \textbf{=} & \multicolumn{1}{r|}{9}          & \multicolumn{1}{r|}{470}        & \multicolumn{1}{r|}{16}         &                                                                                    & \multicolumn{1}{r|}{10}         & \multicolumn{1}{r|}{347}        & \multicolumn{1}{r|}{26}         &                                                                                    & \multicolumn{1}{r|}{11}         & \multicolumn{1}{r|}{302}        & \multicolumn{1}{r|}{30}         &                                                                                    \\ \cline{2-5} \cline{7-9} \cline{11-13}
\multicolumn{1}{|c|}{}                                      & \textbf{-} & \multicolumn{1}{r|}{0}          & \multicolumn{1}{r|}{5}          & \multicolumn{1}{r|}{102}        &                                                                                    & \multicolumn{1}{r|}{0}          & \multicolumn{1}{r|}{15}         & \multicolumn{1}{r|}{132}        &                                                                                    & \multicolumn{1}{r|}{0}          & \multicolumn{1}{r|}{14}         & \multicolumn{1}{r|}{149}        &                                                                                    \\ \hline
\multicolumn{1}{|c|}{\multirow{3}{*}{\textbf{\iffalse Mistral\fi LLM4-FFT}}} & \textbf{+} & \multicolumn{1}{r|}{59}         & \multicolumn{1}{r|}{45}         & \multicolumn{1}{r|}{0}          & \multirow{3}{*}{0.884}                                                              & \multicolumn{1}{r|}{127}        & \multicolumn{1}{r|}{62}         & \multicolumn{1}{r|}{0}          & \multirow{3}{*}{0.837}                                                              & \multicolumn{1}{r|}{163}        & \multicolumn{1}{r|}{55}         & \multicolumn{1}{r|}{0}          & \multirow{3}{*}{\bf 0.854}                                                              \\ \cline{2-5} \cline{7-9} \cline{11-13}
\multicolumn{1}{|c|}{}                                      & \textbf{=} & \multicolumn{1}{r|}{15}         & \multicolumn{1}{r|}{471}        & \multicolumn{1}{r|}{13}         &                                                                                    & \multicolumn{1}{r|}{12}         & \multicolumn{1}{r|}{346}        & \multicolumn{1}{r|}{30}         &                                                                                    & \multicolumn{1}{r|}{8}          & \multicolumn{1}{r|}{293}        & \multicolumn{1}{r|}{22}         &                                                                                    \\ \cline{2-5} \cline{7-9} \cline{11-13}
\multicolumn{1}{|c|}{}                                      & \textbf{-} & \multicolumn{1}{r|}{0}          & \multicolumn{1}{r|}{10}         & \multicolumn{1}{r|}{105}        &                                                                                    & \multicolumn{1}{r|}{0}          & \multicolumn{1}{r|}{13}         & \multicolumn{1}{r|}{128}        &                                                                                    & \multicolumn{1}{r|}{0}          & \multicolumn{1}{r|}{20}         & \multicolumn{1}{r|}{157}        &                                                                                    \\ \hline

\multicolumn{1}{|c|}{\multirow{3}{*}{\textbf{\iffalse Mistral\fi LLM4-LoRA}}} & \textbf{+} & \multicolumn{1}{r|}{63}         & \multicolumn{1}{r|}{53}         & \multicolumn{1}{r|}{0}          & \multirow{3}{*}{0.852}                                                              & \multicolumn{1}{r|}{127}        & \multicolumn{1}{r|}{52}         & \multicolumn{1}{r|}{0}          & \multirow{3}{*}{\bf 0.857}                                                              & \multicolumn{1}{r|}{161}        & \multicolumn{1}{r|}{57}         & \multicolumn{1}{r|}{0}          & \multirow{3}{*}{0.834}                                                              \\ \cline{2-5} \cline{7-9} \cline{11-13}
\multicolumn{1}{|c|}{}                                      & \textbf{=} & \multicolumn{1}{r|}{11}         & \multicolumn{1}{r|}{452}        & \multicolumn{1}{r|}{21}         &                                                                                    & \multicolumn{1}{r|}{12}         & \multicolumn{1}{r|}{356}        & \multicolumn{1}{r|}{26}         &                                                                                    & \multicolumn{1}{r|}{10}         & \multicolumn{1}{r|}{287}        & \multicolumn{1}{r|}{28}         &                                                                                    \\ \cline{2-5} \cline{7-9} \cline{11-13}
\multicolumn{1}{|c|}{}                                      & \textbf{-} & \multicolumn{1}{r|}{0}          & \multicolumn{1}{r|}{21}         & \multicolumn{1}{r|}{97}        &                                                                                    & \multicolumn{1}{r|}{0}          & \multicolumn{1}{r|}{13}         & \multicolumn{1}{r|}{132}        &                                                                                    & \multicolumn{1}{r|}{0}          & \multicolumn{1}{r|}{24}         & \multicolumn{1}{r|}{151}        &                                                                                    \\ \hline

\multicolumn{1}{|c|}{\multirow{3}{*}{\textbf{Human}}}       & \textbf{+} & \multicolumn{1}{r|}{74}         & \multicolumn{1}{r|}{0}          & \multicolumn{1}{r|}{0}          & \multirow{3}{*}{1.000}                                                              & \multicolumn{1}{r|}{139}        & \multicolumn{1}{r|}{0}          & \multicolumn{1}{r|}{0}          & \multirow{3}{*}{1.000}                                                              & \multicolumn{1}{r|}{171}        & \multicolumn{1}{r|}{0}          & \multicolumn{1}{r|}{0}          & \multirow{3}{*}{1.000}                                                              \\ \cline{2-5} \cline{7-9} \cline{11-13}
\multicolumn{1}{|c|}{}                                      & \textbf{=} & \multicolumn{1}{r|}{0}          & \multicolumn{1}{r|}{526}        & \multicolumn{1}{r|}{0}          &                                                                                    & \multicolumn{1}{r|}{0}          & \multicolumn{1}{r|}{421}        & \multicolumn{1}{r|}{0}          &                                                                                    & \multicolumn{1}{r|}{0}          & \multicolumn{1}{r|}{368}        & \multicolumn{1}{r|}{0}          &                                                                                    \\ \cline{2-5} \cline{7-9} \cline{11-13}
\multicolumn{1}{|c|}{}                                      & \textbf{-} & \multicolumn{1}{r|}{0}          & \multicolumn{1}{r|}{0}          & \multicolumn{1}{r|}{118}        &                                                                                    & \multicolumn{1}{r|}{0}          & \multicolumn{1}{r|}{0}          & \multicolumn{1}{r|}{158}        &                                                                                    & \multicolumn{1}{r|}{0}          & \multicolumn{1}{r|}{0}          & \multicolumn{1}{r|}{179}        &                                                                                    \\ \hline
\end{tabular}
\vspace{.05in}\caption{Re-enactment of feature evaluation via QIP relevance, no statistically significant decision on feature launch was negated by the models}\label{tab:feat_rel}
\end{table*}

\subsection{Summary}

To summarize, the findings in the process of model trainings show that 1) including verbose item attribute data can be helpful in training, conditioned on the number of queries per items 2) Full finetuning of the models does not improve over LoRA approximations 3) For the size of the dataset we worked with, $\alpha$s half of rank performed best for LoRA trainings 4) LoRA dropout and learning rate schedulers effectively influence the training and finally, 5) as expected, larger language models stand to benefit more from an augmentation in item textual information (again conditioned on the availability of queries to pair with in the training dataset). We attribute the higher performance results from LoRA compared to the full finetuning to the middle size of the training dataset (about 6 million QIPs) compared to the full model sizes (about 7 billion parameters). A careful balancing of the ratio of model parameters utilized for training to the training data size, is crucial to the outcome.

\section{Relevance Models for Feature Evaluation}
\label{sec:rel}

We discussed the possibility of LLMs replacing humans in evaluating relevance in information retrieval, but how reliable are the models we have trained? To assess the possibility of replacing human evaluation in our feature evaluation tests, which are based on nDCG comparisons of control and variation rankings of items per queries, we employed a number of best models trained in tables \ref{tab:res_llama}, \ref{tab:res_mistral}, and \ref{tab:res_bert} to judge between the control and variation feature automatically, and to compare the LLM feature judgments with that of human evaluators. The results, surprisingly promising, are included below. 

\subsection{LLMs’ Relevance Judgment is Comparable to That of Human Evaluators’}

We have used feature adoption experiments to compare human performance to the LLMs'. From a time-period immediately following the period when all the QIPs used for training and evaluating the LLM models were collected, we extract a set of $718$ experiments in which a paired comparison is performed between a control retrieval ranker and a candidate feature ranker. For each experiment, an independent random sample of queries (from historical customer traffic) was selected to get the top 10 items for each ranker. Query item pairs were then judged by human annotators. The resulting labels were used to calculate the normalized discounted cumulative gain (nDCG)\cite{ndcg} scores for each query and ranker. This was followed by a paired t-test to compare control and feature nDCGs (with $\alpha = 0.05$).

Similarly, we used the LLM models we have trained to judge the same QIPs and calculate the corresponding nDCG scores, followed with statistical significance tests.

In table~\ref{tab:feat_rel}, we show the confusion matrix between the human-based judgments and the LLM based judgments for each model for nDCG@1, nDCG@5 and nDCG@10. Cases where the mean nDCG score for the feature is higher (or lower) than the control with statistical significance ($p<0.05$) are denoted with the symbol $+$ (or $-$). Cases where no statistical significance is observed, regardless of the directionality of the difference, are denoted with symbol $=$. For the ease of comparison across the LLM models, we also compute the ratio of experiments in which the model and human judgments agree on the final outcome (i.e., the sum of the diagonal of each 3x3 subtable, divided by the total count of experiments).

The results of the re-enactment of 718 feature evaluation experiments in table \ref{tab:feat_rel} demonstrate the accuracy of LLMs for relevance evaluation: they agree with human evaluators between 85\% and 89\% of the time in nDCG@10, @5 and @1 respectively, and {\it never} change statistically significant variant evaluations (either favorable or unfavorable) to the opposite. These outcomes suggest finetuned LLMs can be used as reliable first-order relevance evaluators in search and retrieval, hence reducing cost, complexity and length of relevance evaluation, providing the possibility of evaluation at scale. 

To demonstrate scenarios where LLMs can judge relevance better than humans, we have included a few examples in Table (\ref{tab:disagreements}).

\begin{table*}
    \centering
    \begin{tabular}{p{0.15\linewidth} p{0.25\linewidth} p{0.1\linewidth} p{0.1\linewidth} p{0.3\linewidth}}
        Query & Item & Human Label & LLM Pred. & Comment\\
        \hline
        jelly roll vinyl & Sweet Jelly Roll (Vinyl) & Irrelevant & {\bf Relevant} & The item is a song by Jessie Hill. The labeller is apparently confused because they do not have access to the singer attribute.\\
        \hline
        chemical guys tire shine & Chemical Guys Hydrosuds High-Gloss Hyper Foaming SiO2 Ceramic Car Wash Soap & Relevant & {\bf Irrelevant} & This should have been marked/predicted as Related, since the item is used for car wash, but not for tire shine.\\
        \hline
        pioneer 3.5 speakers & Pioneer 6.5 Inch 3-Way 320 Watt Car Coaxial Stereo Speakers Pair & Irrelevant & {\bf Related} & This item is related to the query intent (a larger size than what was requested).\\
    \end{tabular}
    \vspace{.05in}\caption{Where LLMs do better than Hhuamns: Examples of disagreements between test data labels labelled by humans and model predictions, corrected the LLMs' predictions}
    \label{tab:disagreements}
\end{table*}

\section{Discussion: Model Misspecifications}
\label{sec:disc}


To better understand the LLM training outcomes, we analyzed the predictions of the models to understand the cases in which all models mispredicted and the cases in which each individual model made unique mistakes. Additionally, we examined the cases based on query specificity (narrow or broad) and grammar (product type versus brand versus other product attributes). About 125k mispredicted QIPs were used in this analysis. The models, whose predictions are used, are the best models in tables (\ref{tab:res_llama}, \ref{tab:res_mistral}, and \ref{tab:res_bert}).

\subsection{Common Mispredictions among Models} 
Among the unique query-item pairs that were mispredicted by at least one model, 24.6\% were common mispredictions where all models predicted the same incorrect label. When all the models mispredicted, the predicted labels were the same 84\% of the time. Broad and narrow queries of Class-1 labelled items were the most common mispredictions, followed by broad queries of other classes and then narrow queries as recorded in Table \ref{tab:lab_spec}. On the grammar segment, queries that involved product type and attribute intents had the highest frequency of misprediction, while mispredictions in the rest were relatively similar in frequency. The common themes in these mispredictions were 1) ambiguous cases where the mispredicted label was reasonably accurate 2) The presence of spelling errors in the query. 3) handling numerical ranges (size, age, etc.) in the query.

\subsection{Model Specific Mispredictions} 
The models consistently showed similar top mispredicted segments in terms of Product-types, Specificity, and Grammar. LLM4 and LLM3 performed closely, with LLM4 slightly outperforming LLM3. LLM2 showed a performance similar to the two other LLMs when the search query included Product-type intent, but performed worse in other cases. We sampled and studied unique mispredictions from each of the models and have made the following observations: in all segments, the performance of the Walmart pretrained LLM1-Base model consistently lagged behind the other models. While all models made some amount of trivial mispredictions, LLM3 and LLM4 encountered this in cases where the queries were not correctly worded or when dealing with numerical ranges. However, LLM2 and LLM1-base models also showed this issue when attributes were not explicitly stated in the product title or queries. In a few cases, LLM2 and LLM1-base models made mispredictions when the query was verbose while LLM3 and LLM4 did not.  

\begin{table}
    \centering
    \begin{tabular}{l|c}
        Label, specificity & Percentage\\ \hline
        0, broad & 17.5\% \\
        0, narrow & 11.2\%\\
        1, broad & 24.0\%\\
        1. narrow & 18.6\% \\
        2, broad & 15.3\%\\
        2, narrow & 13.3\%\\
    \end{tabular}
    \vspace{.05in}\caption{Common errors by specificity}
    \label{tab:lab_spec}
\end{table}

\section{Conclusion \& Future Directions}
\label{sec:concl}
In this paper, we presented a series of LLM trainings on human-labelled query-item pairs with relevance scores. We presented trade offs and optimizations in data preprocessing, model hyper parameter tuning and model choice. Finally, we used a unique feature evaluation dataset, relevance-labelled by human judges, to demonstrate that large language models can perform on par, at $89\%$ parity at nDCG@1, when employed for feature evaluation automation. 

The models trained here can be used as teacher models for labelling new QIPs for further training of relevance models. The utility of such training model generation is a promising venue for producing synthetic labelled relevance evaluation datasets \cite{yuan23}. Training relevance models on synthetic labels, for comparison with relevance models trained on human labels, is a natural next step.

\small 
\bibliographystyle{ACM-Reference-Format}
\bibliography{references}


\begin{thebibliography}{53}


\ifx \showCODEN    \undefined \def \showCODEN     #1{\unskip}     \fi
\ifx \showDOI      \undefined \def \showDOI       #1{#1}\fi
\ifx \showISBNx    \undefined \def \showISBNx     #1{\unskip}     \fi
\ifx \showISBNxiii \undefined \def \showISBNxiii  #1{\unskip}     \fi
\ifx \showISSN     \undefined \def \showISSN      #1{\unskip}     \fi
\ifx \showLCCN     \undefined \def \showLCCN      #1{\unskip}     \fi
\ifx \shownote     \undefined \def \shownote      #1{#1}          \fi
\ifx \showarticletitle \undefined \def \showarticletitle #1{#1}   \fi
\ifx \showURL      \undefined \def \showURL       {\relax}        \fi
\providecommand\bibfield[2]{#2}
\providecommand\bibinfo[2]{#2}
\providecommand\natexlab[1]{#1}
\providecommand\showeprint[2][]{arXiv:#2}

\bibitem[Abolghasemi et~al\mbox{.}(2022)]%
        {bertimp}
\bibfield{author}{\bibinfo{person}{Amin Abolghasemi}, \bibinfo{person}{Suzan Verberne}, {and} \bibinfo{person}{Leif Azzopard}.} \bibinfo{year}{2022}\natexlab{}.
\newblock \showarticletitle{Improving BERTbased query-by-document retrieval with multi-task optimization}. In \bibinfo{booktitle}{\emph{European Conference on Information Retrieval. Springer}}. \bibinfo{pages}{3--12}.
\newblock


\bibitem[Alonso and Baeza-Yates(2011)]%
        {alonso2011design}
\bibfield{author}{\bibinfo{person}{Omar Alonso} {and} \bibinfo{person}{Ricardo Baeza-Yates}.} \bibinfo{year}{2011}\natexlab{}.
\newblock \showarticletitle{Design and implementation of relevance assessments using crowdsourcing}. In \bibinfo{booktitle}{\emph{European Conference on Information Retrieval}}. Springer, \bibinfo{pages}{153--164}.
\newblock


\bibitem[Alonso et~al\mbox{.}(2009)]%
        {alonso2009can}
\bibfield{author}{\bibinfo{person}{Omar Alonso}, \bibinfo{person}{Stefano Mizzaro}, {et~al\mbox{.}}} \bibinfo{year}{2009}\natexlab{}.
\newblock \showarticletitle{Can we get rid of TREC assessors? Using Mechanical Turk for relevance assessment}. In \bibinfo{booktitle}{\emph{Proceedings of the SIGIR 2009 Workshop on the Future of IR Evaluation}}, Vol.~\bibinfo{volume}{15}. \bibinfo{pages}{16}.
\newblock


\bibitem[Blanco et~al\mbox{.}(2011)]%
        {blanco11}
\bibfield{author}{\bibinfo{person}{Roi Blanco}, \bibinfo{person}{Harry Halpin}, \bibinfo{person}{Daniel~M. Herzig}, \bibinfo{person}{Peter Mika}, \bibinfo{person}{Jeffrey Pound}, \bibinfo{person}{Henry~S. Thompson}, {and} \bibinfo{person}{Duc~Thanh Tran}.} \bibinfo{year}{2011}\natexlab{}.
\newblock \showarticletitle{Repeatable and Reliable Search System Evaluation Using Crowdsourcing}. In \bibinfo{booktitle}{\emph{FSIGIR Conference on Research and Development in Information Retrieval, SIGIR, 2011}}. \bibinfo{pages}{923--932}.
\newblock


\bibitem[Campos et~al\mbox{.}(2023)]%
        {campos2023overview}
\bibfield{author}{\bibinfo{person}{Daniel Campos}, \bibinfo{person}{Surya Kallumadi}, \bibinfo{person}{Corby Rosset}, \bibinfo{person}{Cheng~Xiang Zhai}, {and} \bibinfo{person}{Alessandro Magnani}.} \bibinfo{year}{2023}\natexlab{}.
\newblock \showarticletitle{Overview of the TREC 2023 Product Product Search Track}.
\newblock \bibinfo{journal}{\emph{arXiv preprint arXiv:2311.07861}} (\bibinfo{year}{2023}).
\newblock


\bibitem[Chaudhary et~al\mbox{.}(2023)]%
        {chaudhary23}
\bibfield{author}{\bibinfo{person}{Aditi Chaudhary}, \bibinfo{person}{Karthik Raman}, \bibinfo{person}{Krishna Srinivasan}, \bibinfo{person}{Kazuma Hashimoto}, \bibinfo{person}{Mike Bendersky}, {and} \bibinfo{person}{Marc Najork}.} \bibinfo{year}{2023}\natexlab{}.
\newblock \showarticletitle{Exploring the Viability of Synthetic Query Generation for Relevance Prediction}. In \bibinfo{booktitle}{\emph{ACM SIGIR Workshop on eCommerce (SIGIR eCom 23)}}.
\newblock


\bibitem[Chung et~al\mbox{.}(2022)]%
        {flan22}
\bibfield{author}{\bibinfo{person}{Hyung~Won Chung}, \bibinfo{person}{Le Hou}, \bibinfo{person}{Shayne Longpre}, \bibinfo{person}{Barret Zoph}, \bibinfo{person}{Yi Tay}, \bibinfo{person}{William Fedus}, \bibinfo{person}{Yunxuan Li}, \bibinfo{person}{Xuezhi Wang}, \bibinfo{person}{Mostafa Dehghani}, \bibinfo{person}{Siddhartha Brahma}, \bibinfo{person}{Albert Webson}, \bibinfo{person}{Shixiang~Shane Gu}, \bibinfo{person}{Zhuyun Dai}, \bibinfo{person}{Mirac Suzgun}, \bibinfo{person}{Xinyun Chen}, \bibinfo{person}{Aakanksha Chowdhery}, \bibinfo{person}{Alex Castro-Ros}, \bibinfo{person}{Marie Pellat}, \bibinfo{person}{Kevin Robinson}, \bibinfo{person}{Dasha Valter}, \bibinfo{person}{Sharan Narang}, \bibinfo{person}{Gaurav Mishra}, \bibinfo{person}{Adams Yu}, \bibinfo{person}{Vincent Zhao}, \bibinfo{person}{Yanping Huang}, \bibinfo{person}{Andrew Dai}, \bibinfo{person}{Hongkun Yu}, \bibinfo{person}{Slav Petrov}, \bibinfo{person}{Ed~H. Chi}, \bibinfo{person}{Jeff Dean}, \bibinfo{person}{Jacob Devlin},
  \bibinfo{person}{Adam Roberts}, \bibinfo{person}{Denny Zhou}, \bibinfo{person}{Quoc~V. Le}, {and} \bibinfo{person}{Jason Wei}.} \bibinfo{year}{2022}\natexlab{}.
\newblock \showarticletitle{Scaling Instruction-Finetuned Language Models}.
\newblock \bibinfo{journal}{\emph{arXiv:2210.11416}} (\bibinfo{year}{2022}).
\newblock


\bibitem[Devlin et~al\mbox{.}(2018)]%
        {bert}
\bibfield{author}{\bibinfo{person}{Jacob Devlin}, \bibinfo{person}{Ming{-}Wei Chang}, \bibinfo{person}{Kenton Lee}, {and} \bibinfo{person}{Kristina Toutanova}.} \bibinfo{year}{2018}\natexlab{}.
\newblock \showarticletitle{{BERT:} Pre-training of Deep Bidirectional Transformers for Language Understanding}.
\newblock \bibinfo{journal}{\emph{CoRR}}  \bibinfo{volume}{abs/1810.04805} (\bibinfo{year}{2018}).
\newblock


\bibitem[Faggioli et~al\mbox{.}(2023)]%
        {faggioli23}
\bibfield{author}{\bibinfo{person}{Guglielmo Faggioli}, \bibinfo{person}{Laura Dietz}, \bibinfo{person}{Charles L.~A. Clarke}, \bibinfo{person}{Gianluca Demartini}, \bibinfo{person}{Matthias Hagen}, \bibinfo{person}{Claudia Hauff}, \bibinfo{person}{Noriko Kando}, \bibinfo{person}{Evangelos Kanoulas}, \bibinfo{person}{Martin Potthast}, \bibinfo{person}{Benno Stein}, {and} \bibinfo{person}{Henning Wachsmuth}.} \bibinfo{year}{2023}\natexlab{}.
\newblock \showarticletitle{Perspectives on Large Language Models for Relevance Judgment}. In \bibinfo{booktitle}{\emph{ICTIR, 2023}}. \bibinfo{pages}{39--50}.
\newblock


\bibitem[Falcon and {The PyTorch Lightning team}(2019)]%
        {Falcon_PyTorch_Lightning_2019}
\bibfield{author}{\bibinfo{person}{William Falcon} {and} \bibinfo{person}{{The PyTorch Lightning team}}.} \bibinfo{year}{2019}\natexlab{}.
\newblock \bibinfo{booktitle}{\emph{{PyTorch Lightning}}}.
\newblock
\urldef\tempurl%
\url{https://doi.org/10.5281/zenodo.3828935}
\showDOI{\tempurl}


\bibitem[Gadiraju et~al\mbox{.}(2019)]%
        {gadiraju2019crowd}
\bibfield{author}{\bibinfo{person}{Ujwal Gadiraju}, \bibinfo{person}{Gianluca Demartini}, \bibinfo{person}{Ricardo Kawase}, {and} \bibinfo{person}{Stefan Dietze}.} \bibinfo{year}{2019}\natexlab{}.
\newblock \showarticletitle{Crowd anatomy beyond the good and bad: Behavioral traces for crowd worker modeling and pre-selection}.
\newblock \bibinfo{journal}{\emph{Computer Supported Cooperative Work (CSCW)}}  \bibinfo{volume}{28} (\bibinfo{year}{2019}), \bibinfo{pages}{815--841}.
\newblock


\bibitem[He et~al\mbox{.}(2023a)]%
        {he2023debertav}
\bibfield{author}{\bibinfo{person}{Pengcheng He}, \bibinfo{person}{Jianfeng Gao}, {and} \bibinfo{person}{Weizhu Chen}.} \bibinfo{year}{2023}\natexlab{a}.
\newblock \showarticletitle{De{BERT}aV3: Improving De{BERT}a using {ELECTRA}-Style Pre-Training with Gradient-Disentangled Embedding Sharing}. In \bibinfo{booktitle}{\emph{The Eleventh International Conference on Learning Representations}}.
\newblock
\urldef\tempurl%
\url{https://openreview.net/forum?id=sE7-XhLxHA}
\showURL{%
\tempurl}


\bibitem[He et~al\mbox{.}(2023b)]%
        {he2023que2engage}
\bibfield{author}{\bibinfo{person}{Yunzhong He}, \bibinfo{person}{Yuxin Tian}, \bibinfo{person}{Mengjiao Wang}, \bibinfo{person}{Feier Chen}, \bibinfo{person}{Licheng Yu}, \bibinfo{person}{Maolong Tang}, \bibinfo{person}{Congcong Chen}, \bibinfo{person}{Ning Zhang}, \bibinfo{person}{Bin Kuang}, {and} \bibinfo{person}{Arul Prakash}.} \bibinfo{year}{2023}\natexlab{b}.
\newblock \showarticletitle{Que2Engage: Embedding-based Retrieval for Relevant and Engaging Products at Facebook Marketplace}.
\newblock \bibinfo{journal}{\emph{arXiv preprint arXiv:2302.11052}} (\bibinfo{year}{2023}).
\newblock


\bibitem[Hu et~al\mbox{.}(2022)]%
        {hu2022lora}
\bibfield{author}{\bibinfo{person}{Edward~J Hu}, \bibinfo{person}{yelong shen}, \bibinfo{person}{Phillip Wallis}, \bibinfo{person}{Zeyuan Allen-Zhu}, \bibinfo{person}{Yuanzhi Li}, \bibinfo{person}{Shean Wang}, \bibinfo{person}{Lu Wang}, {and} \bibinfo{person}{Weizhu Chen}.} \bibinfo{year}{2022}\natexlab{}.
\newblock \showarticletitle{Lo{RA}: Low-Rank Adaptation of Large Language Models}. In \bibinfo{booktitle}{\emph{International Conference on Learning Representations}}.
\newblock
\urldef\tempurl%
\url{https://openreview.net/forum?id=nZeVKeeFYf9}
\showURL{%
\tempurl}


\bibitem[Huang et~al\mbox{.}(2013)]%
        {DSSM}
\bibfield{author}{\bibinfo{person}{Po{-}Sen Huang}, \bibinfo{person}{Xiaodong He}, \bibinfo{person}{Jianfeng Gao}, \bibinfo{person}{Li Deng}, \bibinfo{person}{Alex Acero}, {and} \bibinfo{person}{Larry~P. Heck}.} \bibinfo{year}{2013}\natexlab{}.
\newblock \showarticletitle{Learning deep structured semantic models for web search using clickthrough data}. In \bibinfo{booktitle}{\emph{CIKM, 2013}}. \bibinfo{pages}{2333--2338}.
\newblock


\bibitem[J{\"{a}}rvelin and Kek{\"{a}}l{\"{a}}inen(2002)]%
        {ndcg}
\bibfield{author}{\bibinfo{person}{Kalervo J{\"{a}}rvelin} {and} \bibinfo{person}{Jaana Kek{\"{a}}l{\"{a}}inen}.} \bibinfo{year}{2002}\natexlab{}.
\newblock \showarticletitle{Cumulated gain-based evaluation of {IR} techniques}.
\newblock \bibinfo{journal}{\emph{{ACM} Trans. Inf. Syst.}} \bibinfo{volume}{20}, \bibinfo{number}{4} (\bibinfo{year}{2002}), \bibinfo{pages}{422--446}.
\newblock


\bibitem[Jha et~al\mbox{.}(2023)]%
        {jha2023unified}
\bibfield{author}{\bibinfo{person}{Rishikesh Jha}, \bibinfo{person}{Siddharth Subramaniyam}, \bibinfo{person}{Ethan Benjamin}, {and} \bibinfo{person}{Thrivikrama Taula}.} \bibinfo{year}{2023}\natexlab{}.
\newblock \showarticletitle{Unified Embedding Based Personalized Retrieval in Etsy Search}.
\newblock \bibinfo{journal}{\emph{arXiv preprint arXiv:2306.04833}} (\bibinfo{year}{2023}).
\newblock


\bibitem[Jiang et~al\mbox{.}(2023)]%
        {mistral23}
\bibfield{author}{\bibinfo{person}{Albert~Q. Jiang}, \bibinfo{person}{Alexandre Sablayrolles}, \bibinfo{person}{Arthur Mensch}, \bibinfo{person}{Devendra Singh~Chaplot Chris~Bamford}, \bibinfo{person}{Diego de~las Casas}, \bibinfo{person}{Florian Bressand}, \bibinfo{person}{Gianna Lengyel}, \bibinfo{person}{Guillaume Lample}, \bibinfo{person}{Lucile Saulnier}, \bibinfo{person}{Lélio~Renard Lavaud}, \bibinfo{person}{Marie-Anne Lachaux}, \bibinfo{person}{Pierre Stock}, \bibinfo{person}{Teven~Le Scao}, \bibinfo{person}{Thibaut Lavril}, \bibinfo{person}{Thomas Wang}, \bibinfo{person}{Timothée Lacroix}, {and} \bibinfo{person}{William~El Sayed}.} \bibinfo{year}{2023}\natexlab{}.
\newblock \showarticletitle{Mistral 7B}.
\newblock \bibinfo{journal}{\emph{arXiv preprint arXiv:2310.06825}} (\bibinfo{year}{2023}).
\newblock


\bibitem[Jiang et~al\mbox{.}(2019)]%
        {jiang19}
\bibfield{author}{\bibinfo{person}{Yunjiang Jiang}, \bibinfo{person}{Yue Shang}, \bibinfo{person}{Rui Li}, \bibinfo{person}{Wen-Yun Yang}, \bibinfo{person}{Guoyu Tang}, \bibinfo{person}{Chaoyi Ma}, \bibinfo{person}{Yun Xiao}, {and} \bibinfo{person}{Eric Zhao}.} \bibinfo{year}{2019}\natexlab{}.
\newblock \showarticletitle{A unified neural network approach to e-commerce relevance learning}. In \bibinfo{booktitle}{\emph{Proceedings of the 1st International Workshop on Deep Learning Practice for High-Dimensional Sparse Data (DLP-KDD ’19)}}. Association for Computing Machinery, \bibinfo{pages}{1--7}.
\newblock


\bibitem[Kang et~al\mbox{.}(2023)]%
        {kang23llm}
\bibfield{author}{\bibinfo{person}{Wang-Cheng Kang}, \bibinfo{person}{Jianmo Ni}, \bibinfo{person}{Nikhil Mehta}, \bibinfo{person}{Maheswaran Sathiamoorthy}, \bibinfo{person}{Lichan Hong}, \bibinfo{person}{Ed Chi}, {and} \bibinfo{person}{Derek~Zhiyuan Cheng}.} \bibinfo{year}{2023}\natexlab{}.
\newblock \showarticletitle{Do LLMs Understand User Preferences? Evaluating LLMs On User Rating Prediction}.
\newblock \bibinfo{journal}{\emph{arXiv preprint arXiv:2305.06474}} (\bibinfo{year}{2023}).
\newblock


\bibitem[Kazai et~al\mbox{.}(2013)]%
        {kazai2013analysis}
\bibfield{author}{\bibinfo{person}{Gabriella Kazai}, \bibinfo{person}{Jaap Kamps}, {and} \bibinfo{person}{Natasa Milic-Frayling}.} \bibinfo{year}{2013}\natexlab{}.
\newblock \showarticletitle{An analysis of human factors and label accuracy in crowdsourcing relevance judgments}.
\newblock \bibinfo{journal}{\emph{Information retrieval}}  \bibinfo{volume}{16} (\bibinfo{year}{2013}), \bibinfo{pages}{138--178}.
\newblock


\bibitem[Kingma and Ba(2014)]%
        {kingma2014adam}
\bibfield{author}{\bibinfo{person}{Diederik~P Kingma} {and} \bibinfo{person}{Jimmy Ba}.} \bibinfo{year}{2014}\natexlab{}.
\newblock \showarticletitle{Adam: A method for stochastic optimization}.
\newblock \bibinfo{journal}{\emph{arXiv preprint arXiv:1412.6980}} (\bibinfo{year}{2014}).
\newblock


\bibitem[Li et~al\mbox{.}(2021)]%
        {li2021embedding}
\bibfield{author}{\bibinfo{person}{Sen Li}, \bibinfo{person}{Fuyu Lv}, \bibinfo{person}{Taiwei Jin}, \bibinfo{person}{Guli Lin}, \bibinfo{person}{Keping Yang}, \bibinfo{person}{Xiaoyi Zeng}, \bibinfo{person}{Xiao-Ming Wu}, {and} \bibinfo{person}{Qianli Ma}.} \bibinfo{year}{2021}\natexlab{}.
\newblock \showarticletitle{Embedding-based product retrieval in taobao search}. In \bibinfo{booktitle}{\emph{Proceedings of the 27th ACM SIGKDD Conference on Knowledge Discovery \& Data Mining}}. \bibinfo{pages}{3181--3189}.
\newblock


\bibitem[Liu et~al\mbox{.}(2021)]%
        {liu2021que2search}
\bibfield{author}{\bibinfo{person}{Yiqun Liu}, \bibinfo{person}{Kaushik Rangadurai}, \bibinfo{person}{Yunzhong He}, \bibinfo{person}{Siddarth Malreddy}, \bibinfo{person}{Xunlong Gui}, \bibinfo{person}{Xiaoyi Liu}, {and} \bibinfo{person}{Fedor Borisyuk}.} \bibinfo{year}{2021}\natexlab{}.
\newblock \showarticletitle{Que2search: Fast and accurate query and document understanding for search at facebook}. In \bibinfo{booktitle}{\emph{Proceedings of the 27th ACM SIGKDD Conference on Knowledge Discovery \& Data Mining}}. \bibinfo{pages}{3376--3384}.
\newblock


\bibitem[Lu et~al\mbox{.}(2022)]%
        {lu2022ernie}
\bibfield{author}{\bibinfo{person}{Yuxiang Lu}, \bibinfo{person}{Yiding Liu}, \bibinfo{person}{Jiaxiang Liu}, \bibinfo{person}{Yunsheng Shi}, \bibinfo{person}{Zhengjie Huang}, \bibinfo{person}{Shikun Feng~Yu Sun}, \bibinfo{person}{Hao Tian}, \bibinfo{person}{Hua Wu}, \bibinfo{person}{Shuaiqiang Wang}, \bibinfo{person}{Dawei Yin}, {et~al\mbox{.}}} \bibinfo{year}{2022}\natexlab{}.
\newblock \showarticletitle{Ernie-search: Bridging cross-encoder with dual-encoder via self on-the-fly distillation for dense passage retrieval}.
\newblock \bibinfo{journal}{\emph{arXiv preprint arXiv:2205.09153}} (\bibinfo{year}{2022}).
\newblock


\bibitem[Luan et~al\mbox{.}(2021)]%
        {luan2021sparse}
\bibfield{author}{\bibinfo{person}{Yi Luan}, \bibinfo{person}{Jacob Eisenstein}, \bibinfo{person}{Kristina Toutanova}, {and} \bibinfo{person}{Michael Collins}.} \bibinfo{year}{2021}\natexlab{}.
\newblock \showarticletitle{Sparse, Dense, and Attentional Representations for Text Retrieval}.
\newblock \bibinfo{journal}{\emph{Transactions of the Association for Computational Linguistics}}  \bibinfo{volume}{9} (\bibinfo{year}{2021}), \bibinfo{pages}{329--345}.
\newblock


\bibitem[Maddalena et~al\mbox{.}(2016)]%
        {maddalena16}
\bibfield{author}{\bibinfo{person}{Eddy Maddalena}, \bibinfo{person}{Marco Basaldella}, \bibinfo{person}{Dario~De Nart}, \bibinfo{person}{Dante Degl’Innocenti}, \bibinfo{person}{Stefano Mizzaro}, {and} \bibinfo{person}{Gianluca Demartini}.} \bibinfo{year}{2016}\natexlab{}.
\newblock \showarticletitle{Crowdsourcing Relevance Assessments: The Unexpected Benefits of Limiting the Time to Judge}. In \bibinfo{booktitle}{\emph{Fourth AAAI Conference on Human Computation and Crowdsourcing, HCOMP, 2016}}. \bibinfo{pages}{129--138}.
\newblock


\bibitem[Magnani et~al\mbox{.}(2022)]%
        {magnani2022semantic}
\bibfield{author}{\bibinfo{person}{Alessandro Magnani}, \bibinfo{person}{Feng Liu}, \bibinfo{person}{Suthee Chaidaroon}, \bibinfo{person}{Sachin Yadav}, \bibinfo{person}{Praveen Reddy~Suram}, \bibinfo{person}{Ajit Puthenputhussery}, \bibinfo{person}{Sijie Chen}, \bibinfo{person}{Min Xie}, \bibinfo{person}{Anirudh Kashi}, \bibinfo{person}{Tony Lee}, {et~al\mbox{.}}} \bibinfo{year}{2022}\natexlab{}.
\newblock \showarticletitle{Semantic retrieval at walmart}. In \bibinfo{booktitle}{\emph{Proceedings of the 28th ACM SIGKDD Conference on Knowledge Discovery and Data Mining}}. \bibinfo{pages}{3495--3503}.
\newblock


\bibitem[Mangrulkar et~al\mbox{.}(2022)]%
        {peft}
\bibfield{author}{\bibinfo{person}{Sourab Mangrulkar}, \bibinfo{person}{Sylvain Gugger}, \bibinfo{person}{Lysandre Debut}, \bibinfo{person}{Younes Belkada}, \bibinfo{person}{Sayak Paul}, {and} \bibinfo{person}{Benjamin Bossan}.} \bibinfo{year}{2022}\natexlab{}.
\newblock \bibinfo{title}{PEFT: State-of-the-art Parameter-Efficient Fine-Tuning methods}.
\newblock \bibinfo{howpublished}{\url{https://github.com/huggingface/peft}}.
\newblock


\bibitem[Manning et~al\mbox{.}(2008)]%
        {ir}
\bibfield{author}{\bibinfo{person}{Christopher~D. Manning}, \bibinfo{person}{Prabhakar Raghavan}, {and} \bibinfo{person}{Hinrich Sch{\"{u}}tze}.} \bibinfo{year}{2008}\natexlab{}.
\newblock \bibinfo{booktitle}{\emph{Introduction to information retrieval}}.
\newblock \bibinfo{publisher}{Cambridge University Press}.
\newblock
\showISBNx{978-0-521-86571-5}


\bibitem[Mitra and Craswell(2018)]%
        {nir}
\bibfield{author}{\bibinfo{person}{Bhaskar Mitra} {and} \bibinfo{person}{Nick Craswell}.} \bibinfo{year}{2018}\natexlab{}.
\newblock \showarticletitle{An Introduction to Neural Information Retrieval}.
\newblock \bibinfo{journal}{\emph{Foundations and Trends in Information Retrieval}} \bibinfo{volume}{13}, \bibinfo{number}{1} (\bibinfo{year}{2018}), \bibinfo{pages}{1--126}.
\newblock


\bibitem[Muhamed et~al\mbox{.}(2023)]%
        {muhamed2023web}
\bibfield{author}{\bibinfo{person}{Aashiq Muhamed}, \bibinfo{person}{Sriram Srinivasan}, \bibinfo{person}{Choon-Hui Teo}, \bibinfo{person}{Qingjun Cui}, \bibinfo{person}{Belinda Zeng}, \bibinfo{person}{Trishul Chilimbi}, {and} \bibinfo{person}{SVN Vishwanathan}.} \bibinfo{year}{2023}\natexlab{}.
\newblock \showarticletitle{Web-scale semantic product search with large language models}. In \bibinfo{booktitle}{\emph{Pacific-Asia Conference on Knowledge Discovery and Data Mining}}. Springer, \bibinfo{pages}{73--85}.
\newblock


\bibitem[Nigam et~al\mbox{.}(2019)]%
        {nigam2019semantic}
\bibfield{author}{\bibinfo{person}{Priyanka Nigam}, \bibinfo{person}{Yiwei Song}, \bibinfo{person}{Vijai Mohan}, \bibinfo{person}{Vihan Lakshman}, \bibinfo{person}{Weitian Ding}, \bibinfo{person}{Ankit Shingavi}, \bibinfo{person}{Choon~Hui Teo}, \bibinfo{person}{Hao Gu}, {and} \bibinfo{person}{Bing Yin}.} \bibinfo{year}{2019}\natexlab{}.
\newblock \showarticletitle{Semantic product search}. In \bibinfo{booktitle}{\emph{Proceedings of the 25th ACM SIGKDD International Conference on Knowledge Discovery \& Data Mining}}. \bibinfo{pages}{2876--2885}.
\newblock


\bibitem[Nogueira and Cho(2019)]%
        {nogueira2019passage}
\bibfield{author}{\bibinfo{person}{Rodrigo Nogueira} {and} \bibinfo{person}{Kyunghyun Cho}.} \bibinfo{year}{2019}\natexlab{}.
\newblock \showarticletitle{Passage Re-ranking with BERT}.
\newblock \bibinfo{journal}{\emph{arXiv preprint arXiv:1901.04085}} (\bibinfo{year}{2019}).
\newblock


\bibitem[Nouri et~al\mbox{.}(2020)]%
        {nouri2020mining}
\bibfield{author}{\bibinfo{person}{Zahra Nouri}, \bibinfo{person}{Henning Wachsmuth}, {and} \bibinfo{person}{Gregor Engels}.} \bibinfo{year}{2020}\natexlab{}.
\newblock \showarticletitle{Mining crowdsourcing problems from discussion forums of workers}. In \bibinfo{booktitle}{\emph{Proceedings of the 28th International Conference on Computational Linguistics}}. \bibinfo{pages}{6264--6276}.
\newblock


\bibitem[Paszke et~al\mbox{.}(2019)]%
        {NEURIPS2019_9015}
\bibfield{author}{\bibinfo{person}{Adam Paszke}, \bibinfo{person}{Sam Gross}, \bibinfo{person}{Francisco Massa}, \bibinfo{person}{Adam Lerer}, \bibinfo{person}{James Bradbury}, \bibinfo{person}{Gregory Chanan}, \bibinfo{person}{Trevor Killeen}, \bibinfo{person}{Zeming Lin}, \bibinfo{person}{Natalia Gimelshein}, \bibinfo{person}{Luca Antiga}, \bibinfo{person}{Alban Desmaison}, \bibinfo{person}{Andreas Kopf}, \bibinfo{person}{Edward Yang}, \bibinfo{person}{Zachary DeVito}, \bibinfo{person}{Martin Raison}, \bibinfo{person}{Alykhan Tejani}, \bibinfo{person}{Sasank Chilamkurthy}, \bibinfo{person}{Benoit Steiner}, \bibinfo{person}{Lu Fang}, \bibinfo{person}{Junjie Bai}, {and} \bibinfo{person}{Soumith Chintala}.} \bibinfo{year}{2019}\natexlab{}.
\newblock \showarticletitle{PyTorch: An Imperative Style, High-Performance Deep Learning Library}.
\newblock In \bibinfo{booktitle}{\emph{Advances in Neural Information Processing Systems 32}}, \bibfield{editor}{\bibinfo{person}{H.~Wallach}, \bibinfo{person}{H.~Larochelle}, \bibinfo{person}{A.~Beygelzimer}, \bibinfo{person}{F.~d\textquotesingle Alch\'{e}-Buc}, \bibinfo{person}{E.~Fox}, {and} \bibinfo{person}{R.~Garnett}} (Eds.). \bibinfo{publisher}{Curran Associates, Inc.}, \bibinfo{pages}{8024--8035}.
\newblock
\urldef\tempurl%
\url{http://papers.neurips.cc/paper/9015-pytorch-an-imperative-style-high-performance-deep-learning-library.pdf}
\showURL{%
\tempurl}


\bibitem[Paul~Thomas and Mitra(2023)]%
        {thomas2023}
\bibfield{author}{\bibinfo{person}{Nick~Craswell Paul~Thomas, Seth~Spielman} {and} \bibinfo{person}{Bhaskar Mitra}.} \bibinfo{year}{2023}\natexlab{}.
\newblock \showarticletitle{Large language models can accurately predict searcher preferences}.
\newblock \bibinfo{journal}{\emph{arXiv:2309.10621}} (\bibinfo{year}{2023}).
\newblock


\bibitem[Qu et~al\mbox{.}(2021)]%
        {qu2021rocketqa}
\bibfield{author}{\bibinfo{person}{Yingqi Qu}, \bibinfo{person}{Yuchen Ding}, \bibinfo{person}{Jing Liu}, \bibinfo{person}{Kai Liu}, \bibinfo{person}{Ruiyang Ren}, \bibinfo{person}{Wayne~Xin Zhao}, \bibinfo{person}{Daxiang Dong}, \bibinfo{person}{Hua Wu}, {and} \bibinfo{person}{Haifeng Wang}.} \bibinfo{year}{2021}\natexlab{}.
\newblock \showarticletitle{RocketQA: An optimized training approach to dense passage retrieval for open-domain question answering}. In \bibinfo{booktitle}{\emph{NAACL-HLT}}.
\newblock


\bibitem[Reddy et~al\mbox{.}(2022)]%
        {reddy2022shopping}
\bibfield{author}{\bibinfo{person}{Chandan~K. Reddy}, \bibinfo{person}{Lluís Màrquez}, \bibinfo{person}{Fran Valero}, \bibinfo{person}{Nikhil Rao}, \bibinfo{person}{Hugo Zaragoza}, \bibinfo{person}{Sambaran Bandyopadhyay}, \bibinfo{person}{Arnab Biswas}, \bibinfo{person}{Anlu Xing}, {and} \bibinfo{person}{Karthik Subbian}.} \bibinfo{year}{2022}\natexlab{}.
\newblock \bibinfo{title}{Shopping Queries Dataset: A Large-Scale {ESCI} Benchmark for Improving Product Search}.
\newblock
\newblock
\showeprint[arxiv]{2206.06588}


\bibitem[Reimers and Gurevych(2019)]%
        {reimers19}
\bibfield{author}{\bibinfo{person}{Nils Reimers} {and} \bibinfo{person}{Iryna Gurevych}.} \bibinfo{year}{2019}\natexlab{}.
\newblock \showarticletitle{Sentence-BERT: Sentence Embeddings using Siamese BERT-Networks}. In \bibinfo{booktitle}{\emph{Proceedings of the 2019 Conference on Empirical Methods in Natural Language Processing and the 9th International Joint Conference on Natural Language Processing (EMNLP-IJCNLP)}}. \bibinfo{pages}{3982--3992}.
\newblock


\bibitem[Shan et~al\mbox{.}(2023)]%
        {shan23}
\bibfield{author}{\bibinfo{person}{Hongyu Shan}, \bibinfo{person}{Qishen Zhang}, \bibinfo{person}{Zhongyi Liu}, \bibinfo{person}{Guannan Zhang}, {and} \bibinfo{person}{Chenliang Li}.} \bibinfo{year}{2023}\natexlab{}.
\newblock \showarticletitle{Beyond Two-Tower: Attribute Guided Representation Learning for Candidate Retrieval}. In \bibinfo{booktitle}{\emph{WWW '23: Proceedings of the ACM Web Conference 2023}}. \bibinfo{pages}{3173--3178}.
\newblock


\bibitem[Sheshadri and Lease(2013)]%
        {sheshadri2013square}
\bibfield{author}{\bibinfo{person}{Aashish Sheshadri} {and} \bibinfo{person}{Matthew Lease}.} \bibinfo{year}{2013}\natexlab{}.
\newblock \showarticletitle{Square: A benchmark for research on computing crowd consensus}. In \bibinfo{booktitle}{\emph{Proceedings of the AAAI Conference on Human Computation and Crowdsourcing}}, Vol.~\bibinfo{volume}{1}. \bibinfo{pages}{156--164}.
\newblock


\bibitem[Srivastava et~al\mbox{.}(2014)]%
        {dropout}
\bibfield{author}{\bibinfo{person}{Nitish Srivastava}, \bibinfo{person}{Geoffrey~E. Hinton}, \bibinfo{person}{Alex Krizhevsky}, \bibinfo{person}{Ilya Sutskever}, {and} \bibinfo{person}{Ruslan Salakhutdinov}.} \bibinfo{year}{2014}\natexlab{}.
\newblock \showarticletitle{Dropout: a simple way to prevent neural networks from overfitting}.
\newblock \bibinfo{journal}{\emph{Journal of Machine Learning Research}} \bibinfo{volume}{15}, \bibinfo{number}{1} (\bibinfo{year}{2014}), \bibinfo{pages}{1929--1958}.
\newblock


\bibitem[Sun et~al\mbox{.}(2023)]%
        {sun23}
\bibfield{author}{\bibinfo{person}{Xiaojie Sun}, \bibinfo{person}{Keping Bi}, \bibinfo{person}{Jiafeng Guo}, \bibinfo{person}{Xinyu Ma}, \bibinfo{person}{Yixing Fan}, \bibinfo{person}{Hongyu Shan}, \bibinfo{person}{Qishen Zhang}, {and} \bibinfo{person}{Zhongyi Liu}.} \bibinfo{year}{2023}\natexlab{}.
\newblock \showarticletitle{Pre-training with Aspect-Content Text Mutual Prediction for Multi-Aspect Dense Retrieval}. In \bibinfo{booktitle}{\emph{CIKM '23: Proceedings of the 32nd ACM International Conference on Information and Knowledge Management}}. \bibinfo{pages}{4300--4304}.
\newblock


\bibitem[Tamine and Chouquet(2017)]%
        {tamine2017impact}
\bibfield{author}{\bibinfo{person}{Lynda Tamine} {and} \bibinfo{person}{Cecile Chouquet}.} \bibinfo{year}{2017}\natexlab{}.
\newblock \showarticletitle{On the impact of domain expertise on query formulation, relevance assessment and retrieval performance in clinical settings}.
\newblock \bibinfo{journal}{\emph{Information Processing \& Management}} \bibinfo{volume}{53}, \bibinfo{number}{2} (\bibinfo{year}{2017}), \bibinfo{pages}{332--350}.
\newblock


\bibitem[Touvron et~al\mbox{.}(2023)]%
        {llama23}
\bibfield{author}{\bibinfo{person}{Hugo Touvron}, \bibinfo{person}{Thibaut Lavril}, \bibinfo{person}{Gautier Izacard}, \bibinfo{person}{Xavier Martinet}, \bibinfo{person}{Marie-Anne Lachaux}, \bibinfo{person}{Timothée Lacroix}, \bibinfo{person}{Baptiste Rozière}, \bibinfo{person}{Naman Goyal}, \bibinfo{person}{Eric Hambro}, \bibinfo{person}{Faisal Azhar}, \bibinfo{person}{Aurelien Rodriguez}, \bibinfo{person}{Armand Joulin}, \bibinfo{person}{Edouard Grave}, {and} \bibinfo{person}{Guillaume Lample}.} \bibinfo{year}{2023}\natexlab{}.
\newblock \showarticletitle{LLaMA: Open and Efficient Foundation Language Models}.
\newblock \bibinfo{journal}{\emph{arXiv preprint arXiv:2302.13971}} (\bibinfo{year}{2023}).
\newblock


\bibitem[Vaswani et~al\mbox{.}(2017)]%
        {transformer}
\bibfield{author}{\bibinfo{person}{Ashish Vaswani}, \bibinfo{person}{Noam Shazeer}, \bibinfo{person}{Niki Parmar}, \bibinfo{person}{Jakob Uszkoreit}, \bibinfo{person}{Llion Jones}, \bibinfo{person}{Aidan~N. Gomez}, \bibinfo{person}{Lukasz Kaiser}, {and} \bibinfo{person}{Illia Polosukhin}.} \bibinfo{year}{2017}\natexlab{}.
\newblock \showarticletitle{Attention is All you Need}. In \bibinfo{booktitle}{\emph{NeurIPS}}. \bibinfo{pages}{6000--6010}.
\newblock


\bibitem[Wolf et~al\mbox{.}(2020)]%
        {wolf2020huggingfaces}
\bibfield{author}{\bibinfo{person}{Thomas Wolf}, \bibinfo{person}{Lysandre Debut}, \bibinfo{person}{Victor Sanh}, \bibinfo{person}{Julien Chaumond}, \bibinfo{person}{Clement Delangue}, \bibinfo{person}{Anthony Moi}, \bibinfo{person}{Pierric Cistac}, \bibinfo{person}{Tim Rault}, \bibinfo{person}{Rémi Louf}, \bibinfo{person}{Morgan Funtowicz}, \bibinfo{person}{Joe Davison}, \bibinfo{person}{Sam Shleifer}, \bibinfo{person}{Patrick von Platen}, \bibinfo{person}{Clara Ma}, \bibinfo{person}{Yacine Jernite}, \bibinfo{person}{Julien Plu}, \bibinfo{person}{Canwen Xu}, \bibinfo{person}{Teven~Le Scao}, \bibinfo{person}{Sylvain Gugger}, \bibinfo{person}{Mariama Drame}, \bibinfo{person}{Quentin Lhoest}, {and} \bibinfo{person}{Alexander~M. Rush}.} \bibinfo{year}{2020}\natexlab{}.
\newblock \bibinfo{title}{HuggingFace's Transformers: State-of-the-art Natural Language Processing}.
\newblock
\newblock
\showeprint[arxiv]{1910.03771}~[cs.CL]


\bibitem[Xu et~al\mbox{.}(2020)]%
        {xuy20}
\bibfield{author}{\bibinfo{person}{Yige Xu}, \bibinfo{person}{Xipeng Qiu}, \bibinfo{person}{Ligao Zhou}, {and} \bibinfo{person}{Xuanjing Huang}.} \bibinfo{year}{2020}\natexlab{}.
\newblock \showarticletitle{Improving bert fine-tuning via self-ensemble and self-distillation}.
\newblock \bibinfo{journal}{\emph{arXiv preprint arXiv:2002.10345}} (\bibinfo{year}{2020}).
\newblock


\bibitem[Yang et~al\mbox{.}(2019)]%
        {yang2019simple}
\bibfield{author}{\bibinfo{person}{Wei Yang}, \bibinfo{person}{Haotian Zhang}, {and} \bibinfo{person}{Jimmy Lin}.} \bibinfo{year}{2019}\natexlab{}.
\newblock \showarticletitle{Simple applications of BERT for ad hoc document retrieval}.
\newblock \bibinfo{journal}{\emph{arXiv preprint arXiv:1903.10972}} (\bibinfo{year}{2019}).
\newblock


\bibitem[Yao et~al\mbox{.}(2021)]%
        {yao2021learning}
\bibfield{author}{\bibinfo{person}{Shaowei Yao}, \bibinfo{person}{Jiwei Tan}, \bibinfo{person}{Xi Chen}, \bibinfo{person}{Keping Yang}, \bibinfo{person}{Rong Xiao}, \bibinfo{person}{Hongbo Deng}, {and} \bibinfo{person}{Xiaojun Wan}.} \bibinfo{year}{2021}\natexlab{}.
\newblock \showarticletitle{Learning a product relevance model from click-through data in e-commerce}. In \bibinfo{booktitle}{\emph{Proceedings of the Web Conference 2021}}. \bibinfo{pages}{2890--2899}.
\newblock


\bibitem[Yuan et~al\mbox{.}(2023)]%
        {yuan23}
\bibfield{author}{\bibinfo{person}{Weizhe Yuan}, \bibinfo{person}{Richard~Yuanzhe Pang}, \bibinfo{person}{Kyunghyun Cho}, \bibinfo{person}{Sainbayar Sukhbaatar}, \bibinfo{person}{Jing Xu}, {and} \bibinfo{person}{Jason Weston}.} \bibinfo{year}{2023}\natexlab{}.
\newblock \showarticletitle{Self-Rewarding Language Models}.
\newblock \bibinfo{journal}{\emph{arXiv preprint arXiv:2401.10020}} (\bibinfo{year}{2023}).
\newblock


\bibitem[Zhang et~al\mbox{.}(2020)]%
        {zhang2020towards}
\bibfield{author}{\bibinfo{person}{Han Zhang}, \bibinfo{person}{Songlin Wang}, \bibinfo{person}{Kang Zhang}, \bibinfo{person}{Zhiling Tang}, \bibinfo{person}{Yunjiang Jiang}, \bibinfo{person}{Yun Xiao}, \bibinfo{person}{Weipeng Yan}, {and} \bibinfo{person}{Wen-Yun Yang}.} \bibinfo{year}{2020}\natexlab{}.
\newblock \showarticletitle{Towards personalized and semantic retrieval: An end-to-end solution for e-commerce search via embedding learning}. In \bibinfo{booktitle}{\emph{Proceedings of the 43rd International ACM SIGIR Conference on Research and Development in Information Retrieval}}. \bibinfo{pages}{2407--2416}.
\newblock


\end{thebibliography}

\end{document}